\documentclass[fleqn,usenatbib]{mnras}

\usepackage{newtxtext,newtxmath}
\usepackage[T1]{fontenc}

\DeclareRobustCommand{\VAN}[3]{#2}
\let\VANthebibliography\thebibliography
\def\thebibliography{\DeclareRobustCommand{\VAN}[3]{##3}\VANthebibliography}

\usepackage{graphicx}	% Including figure files
\usepackage{amsmath}	% Advanced maths commands
\usepackage{longtable}
\usepackage{pdflscape}
\usepackage{textgreek}

\title[Lunar Impact Flash observations of the 2025 Geminids]{Twin Impact Lunar Telescope network: Lunar Impact Flash observations of the 2025 Geminids}

\author[D. Sheward et al]{Daniel Sheward$^{1}$,\thanks{E-mail: ds869@leicester.ac.uk}
Marco Delbo$^{2,1}$,
Chrysa Avdellidou$^{1}$,
Philippe Lognonné$^{3}$,
Jeremie Vaubaillon$^{4}$,
\newauthor
Pierre-Yves Froissart$^{3}$,
Christelle Saliby$^{5}$,
Paul Girard$^{5}$, 
Nicolas Mauclert$^{5}$,
Bruno Mongellaz$^{5}$,
\newauthor
Laurent Herrier$^{5}$,
Nicolas Anfosso$^{5}$,
Thierry Parra$^{5}$,
Fausto Giacometti$^{6}$,
Andrea Ferrero$^{7}$,
\newauthor
Jean-Pierre Rivet$^{2}$,
Elisa Maria Alessi$^{8}$,
Anthony Cook$^{9}$,
Detlef Koschny$^{10}$,
James Dawson$^{11}$,
Chris Hooker$^{11}$,
\newauthor
Alex Pratt$^{11}$,
Michael O'Connell$^{12}$,
Aldo Tonon$^{13}$,
Vincenzo della Vecchia$^{13}$,
Luigi Zanatta$^{13}$
\\
$^{1}$University of Leicester, School of Physics and Astronomy, University Road, LE1 7RH, Leicester, UK.\\
$^{2}$Université Côte d’Azur, CNRS–Lagrange, Observatoire de la Côte d’Azur, CS 34229, F 06304 NICE Cedex 4, France.\\
$^{3}$Université Paris Cité, Institut de Physique du Globe de Paris, CNRS, Paris, France.\\
$^{4}$ LTE, Observatoire de Paris, Université PSL, Sorbonne Université, Université de Lille, LNE, CNRS, 61 Avenue de l’Observatoire, Paris 75014, France.\\
$^{5}$Observatoire de la Côte d’Azur, CS 34229, F 06304 NICE Cedex 4, France.\\
$^{6}$F.\&F. Giacometti, Viale della pace n.104. 45100 Rovigo
Italia.\\
$^{7}$Bigmuskie Observatory (B88), via Italo Aresca 12, 14047 Moberelli, Asti, Italy.\\
$^{8}$Istituto di Matematica Applicata e Tecnologie Informatiche, Consiglio Nazionale delle Ricerche, via Alfonso Corti 12, 20133 Milan, Italy.\\
$^{9}$Aberystwyth University, Department of Physics, Aberystwyth SY23 3FL, UK.\\
$^{10}$Lunar and Planetary Exploration, Technical University of Munich Lise-Meitner-Str. 9, D-85521 Ottobrunn.\\
$^{11}$British Astronomical Association, PO Box 702, Tonbridge TN9 9TX\\
$^{12}$R44 Metis Trevinca Skies, A. Veiga, 32360, Ourense, Spain\\
$^{13}$Unione Astrofili Italiani, Italy}
\date{Accepted XXX. Received YYY; in original form ZZZ}

\pubyear{\the\year{}}

\begin{document}
\label{firstpage}
\pagerange{\pageref{firstpage}--\pageref{lastpage}}
\maketitle

\begin{abstract}
Meteoroid impacts on the Moon, observed from Earth as flashes typically lasting a few tens of milliseconds, have been monitored for three decades for determining meteoroids’ size and mass frequency distribution in the cm-dm range. Studies link these observed impact events to fresh craters advancing our understanding of energy partitioning during an impact. Currently we are transitioning to a new era where lunar impact flashes (LIFs) can be used to supplement upcoming lunar seismology to study the internal lunar structure. Here we present results from the first station of a telescope network under development for continuous LIF monitoring. Observations were carried out during the Geminids 2025 campaign, initiated by the LUMIO Science Team in the framework of their public engagement activities. We detected 53 potential impact flashes and confirmed 11 of them through multi-frame observations, and independent detections by other observers. We present evidence suggesting that some of the yet-unconfirmed events may be real. Our confirmed events range between magnitude +7.5 and +10.4, primarily in the V- and R-band. We obtained a high rate of observations per hour, highlighting the importance of high-ZHR meteoroid streams for observing LIFs. We also discuss the scientific value of potential LIFs that remain unconfirmed in optical data alone. Even without multi-station confirmation, these events can correlate to seismic signals in future lunar seismic networks, thereby providing useful physical constraints on impact processes. This approach would also allow stations equipped with a single telescope/camera to meaningfully contribute to the network.

\end{abstract}


%%%%%%%%%%%%%%%%%%%%%%%%%%%%%%%%%%%%%%%%%%%%%%%%%%

%%%%%%%%%%%%%%%%% BODY OF PAPER %%%%%%%%%%%%%%%%%%

\section{Introduction}
Since late 1990s, the lunar surface has been monitored in order to detect the transient light phenomena that occur during meteoroid hypervelocity impact events, called lunar impact flashes (LIF)~\citep{ortiz1999, ortiz2000, yanagisawa2002, suggs2014}. Ground-based observations of LIFs provide a unique means to monitor the present-day flux of meteoroids striking the Moon, quantify the timing and location of the impacts and their corresponding luminous energies released upon impact, the latter being a small amount of the initial kinetic energy of the meteoroid \citep{bellotrubio2000A, swift2011, bouley2012,suggs2014,avdellidou2021,liakos2024}. Impact flashes result from the rapid heating and vaporization of meteoroid material and regolith \citep[average temperatures $\sim$2800\,K, see][]{madiedo2018,avdellidou2019,madiedo2019, avdellidou2021, yanagisawa2021, yanagisawa2022} during hypervelocity impacts. Measured LIFs have, in general, a duration between few tens (typical) and several hundreds of milliseconds for the brightest events \citep[][ and references therein]{avdellidou2021}, with some exceptionally energetic events having a duration in the order of seconds~\citep{suggs2014}, having been observed as long as 8 seconds~\citep{madiedo2014}. To date, more than 850 LIFs have been reported in the literature or archived in public databases \citep{suggs2014, avdellidou2021,sheward2025}. Yet, LIFs are rare events with rates of about one every few hours \citep{suggs2014,liakos2024} for those with magnitudes $<$11 in the V band, which are therefore observable with telescopes with diameters between 0.35\,m \citep{suggs2014} and 1.2\,m \citep{xilouris2018}. This rate can increase when meteoroid streams hit the Moon, while sporadic meteoroids (those not obviously related to meteoroid streams) hit the Moon at a rate lower than the average value \citep{liakos2024}.

The analysis of LIFs and their link to meteoroid streams returns the mass and size of the impacting meteoroids, since their impact speed is known for stream meteoroids \citep{suggs2014,avdellidou2019,avdellidou2021}, while subsequent search in lunar orbiter data allows the discovery of the fresh craters. The latter has been done successfully using Lunar Reconnaissance Orbiter data \citep{suggs2014,sheward2022,sheward2025}. The link of a meteoroid with specific physical properties to a fresh crater with measured size is the fundamental step in studying impact scaling laws with applications, among others, in planetary surface chronology.

Impacts of meteoroids can also generate seismic waves within the Moon that can be measured by seismometers placed on the lunar surface \citep{Chenet2006,Lognonne2003E&PSL.211...27L}. These measurements allow one to determine the vertical component of the momentum of the impactor, as demonstrated in the case of the planet Mars using the seismometer on board NASA's InSight mission \citep{Posiolova2022Sci...378..412P}. The thickness of the lunar crust, and nature of the upper mantle, between the seismometer and impact points can be inferred from analysis of the seismic events for those impacts which can be geolocated and timed by their LIFs. 

Using data from the Apollo's Passive Seismic Experiments \cite{lognonne2009} showed that the rate of impact seismic detection on the Moon is in agreement with those in Earth's atmosphere, as detected by military early warning satellites \citep{brown2002} for large impacts, or, for small ones, by meteor observations \citep{revelle2001} or lunar impact flash observations in visible light \citep{suggs2014,avdellidou2021}. These results provide the perspective for new seismic missions on the Moon. 

Following the success of the InSight Mission, %where a seismometer was deployed on Mars to listen to Marsquakes and Mars impacts, 
we are entering a renaissance-era for lunar seismology: In the coming years there are currently four seismometers scheduled to go to the Moon. The \hbox{CNSAs} Chang'e-7 mission will deploy a seismometer near the South pole, and is scheduled to launch August 2026~\citep{wang2024}. The Farside Seismic Suite (FSS) is planned to be launched in 2027 to Schr\"odinger Basin as part of NASA's Commercial Lunar Program Services program~\citep{panning2025abs}. The Lunar Environment Monitoring Station (LEMS) is scheduled to be deployed at the lunar South pole by NASA astronauts on board the Artemis III mission (currently targeted for 2027-2028)~\citep{schmerr2024}. The South Pole Seismic Station (SPSS) is planned to be deployed at the lunar South pole by NASA's Artemis IV astronauts (currently targeted for 2028/2029)~\citep{artemis4pressrelease}. These instruments have the goal of detecting deep moonquakes, shallow moonquakes, and the seismic signals generated by impacts, to investigate the interior structure of the Moon. In this context, analysis of impacts are crucial for the probing of the lunar interior, as the formed crater can be detected through surface imaging \citep[e.g.,][]{sheward2025}, providing a highly constrained epicentral location for the seismic source. 

Locating the post-impact crater does not provide full constraints for the seismic source, notably the time of the impact, which is needed to directly measure the travel time of the seismic signal between the impact location and the seismometer. The proximity of the Moon enables multi-messenger observations to occur: namely, by simultaneously observing for LIFs while also listening with in-situ seismometers, one can obtain the luminous energy of the flash (and therefore the kinetic energy), the precise time of the LIF, and its location on the Moon from the LIF observation, and subsequently observe the waveform measured by the seismometer. This leaves the lunar interior as the only source of uncertainty, allowing for high quality inversion of the thickness of the lunar crust to be performed.

It is clear that LIFs that are brighter and closer to the seismometer are the most likely to generate measurable seismic signals with high signal-to-noise ratio. While larger impactors produce stronger seismic signals, they are correspondingly rarer on the Moon, and this rarity is further increased by the requirement that they occur in  proximity to the seismometer. 

For the reasons discussed above, sustained continuous monitoring of LIFs is necessary. To the best of our knowledge, currently, two professional surveys dedicated to LIF detection are active. The ESA-funded NELIOTA programme \citep{xilouris2018,bonanos2018}, which uses the 1.2-m Kryoneri Observatory telescope to monitor the non-illuminated lunar hemisphere in visible light. NELIOTA typically observes around ten nights per month, for a few hours after sunset following new Moon, and for a few hours before sunrise preceding new Moon. NASA's Meteoroid Environment Office has also restarted their LIF observations, however only observe at periods of high stream activity, and have not made any observations publicly available. 

In order to increase LIF monitoring time, we are developing at least three LIF observing stations located in France, California (US) and Australia. These stations are based on the Twin Impact Lunar Telescope (TILT) concept. This consists of a twin, 40\,cm diameter, 1600-1700\,mm focal length, co-aligned telescope system optimized for high-cadence lunar observations in the near-infrared, where the thermal emission of typical LIFs peaks as derived from  their temperature distribution \citep{avdellidou2021}. Operating in the infrared provides a major advantage: stray light from the illuminated portion of the Moon is significantly reduced, as reflected sunlight by the Moon peaks in the visible light ($\sim$0.5\,µm), and daytime/twilight sky brightness is strongly attenuated because Rayleigh scattering decreases with wavelength $\lambda$$^{-4}$. These effects enable more efficient observations close to the illuminated limb and even allow daytime observations \citep{sheward2024}. The twin configuration further discriminates genuine flashes from false-positives, while the deployment of three geographically distributed stations ensures extended temporal coverage and mitigates weather-related losses. The TILT network will provide well-timed and localized impact events that can be combined with future lunar seismic measurements, supplying natural impact sources of known energy and location for crustal characterization.

During the 2025 Geminids, TILT1 (France) participated in a global campaign to observe LIFs. This campaign was initiated by the Science Team of the ESA LUMIO cubesat \citep{cervone2022,topputo2023} in the framework of their public engagement activities~\citep[in prep.]{koschny2026}. 

As a high zenithal hourly rate (ZHR) meteoroid stream, the Geminids have been utilised multiple times throughout literature for the purposes of LIF science. \cite{yanagisawa2008} first reported five LIFs observed during the 2007 Geminids, and later in 2018, observed the Geminids to obtain low dispersion spectra of thirteen LIFs~\citep{yanagisawa2021}. \cite{ortiz2015} observed during the 2007, 2011, 2013, and 2014 Geminids, and detected 12 lunar impact flashes over their four observing sessions, nine of which could be linked to the Geminid meteoroid stream. These impact flashes were estimated to have a luminous efficiency, $\eta$ of around $2.1\times10^{-3}$ through comparison of the observed impacts to the expected distribution of Geminids, following the methodology set out in \cite{bellotrubio2000, bellotrubio2000A}.

In the following, we present the instrument configuration, the mode of observations, LIF detection and data analysis, as well as their connection to potential lunar seismic activity. 

\section{TILT Instrument}
The TILT network of ground-based optical instruments 
%specifically designed for the high–speed detection and characterization of meteoroid impacts on the lunar surface. The system 
is funded through the ERC Advanced Grant LISTEN-FLASH. The first station, referred to as TILT1 (or simply TILT), was installed in summer 2025 at the Calern Observatory, within the Observatoire de la Côte d’Azur (OCA), France. 
%Each station consists of two 400-mm in aperture telescopes for simultaneous detection of LIFs.

TILT1 employs a twin classical Newtonian configuration using a parabolic primary mirror with a focal length of 1700\,mm, and aperture of 400\,mm. The next units, TILT2 and TILT3 (planned for deployment in the United States and Australia, respectively) will adopt a Newtonian hyperbolic astrograph architecture with a Ross/Rosin corrector, optimized for wide-field imaging of the lunar nearside. These systems will have a focal length of 1600\,mm, and aperture of 400\,mm. This optical configuration combines fast throughput with a moderate plate scale, providing both sensitivity to faint (sub-second) optical flashes and adequate spatial sampling for localization and coincidence checks between units.

Precise tracking of the lunar disk is essential for continuous high-speed imaging. TILT stations employ custom alt–azimuth fork mounts engineered to support the 400-mm optical assemblies. The choice of an alt–az configuration reduces mechanical footprint and avoids meridian flips that would otherwise interrupt data acquisition. The mounts use friction-drive actuation, in which a motor-driven roller is pressed against a large disk coupled to each axis. High-resolution absolute encoders (Heidenhain) provide closed-loop pointing and tracking. Each optical tube is equipped with a field rotator, allowing both detector orientation with respect to the lunar terminator and compensation for field rotation during alt–az tracking.

\section{Observations}
TILTs are conceived to operate primarily in the near-infrared wavelengths (NIR; also referred to as the short-wave infrared, SWIR). However, during the 2025 Geminids observations we also utilised visible-light CMOS cameras. On 13 December, we operated a ZWO ASI183MM and a ZWO ASI174MM mounted on the TILT tube~1 and tube~2, respectively, and observed between 01:30 and 06:00 UTC. On 14 December, the ZWO ASI183MM remained on tube~1, while a Raptor Photonics Ninox~640\,SU was installed on tube~2 and we observed between 02:30-06:30 UTC. All cameras on both nights were operated without optical filters.

The ZWO ASI183MM, ZWO ASI174MM, and Ninox~640\,SU were all operated with an exposure time of 50\,ms, however due to the differing read-out times of the three cameras, we obtained nominal frame rates of approximately $20\,\mathrm{fps}$, $19\,\mathrm{fps}$, and $17\,\mathrm{fps}$ respectively. The two ASI Cameras were operated with a gain of 0.1\,e$^-$/ADU, and the Ninox in low gain mode - which gives a gain of 0.9-2.5\,e$^-$/ADU.

In our observational configuration, the typical readout period of the cameras is around 2\,msec for the two ASI cameras, and 10\,msec for the Ninox. It was discovered after observations had concluded that, due to a computer configuration issue, the time between subsequent frames could be vary by tens to hundreds of milliseconds. Consequently, with an exposure time of 50\,msec, and often 50-100\,msec between subsequent exposures, for a majority of the observed time our system was not capturing frames. Thus, while the cameras' software were reporting the aforementioned frame rates, the actual frame rates obtained (stored on computer disks) varied over the course of observation. This would undoubtedly cause a considerable number of LIF candidates to be missed, a fact which is backed up by the many LIF candidates reported by other observers that are not in our data. This, coupled with the second camera's lower sensitivity, meant that for these observations, we were unable to self-confirm single frame flashes.

In addition, while the Ninox~640\,SU can be externally triggered, the ZWO ASI183MM and ASI174MM cameras cannot. Consequently, during both nights of observations the exposures of the cameras on tubes~1 and~2 of TILT were not synchronized. The limiting magnitudes for each of the three cameras we used during these observations are estimated to be V$\approx$11.9 for the ASI cameras, and J$\approx$11.1 for the Ninox.

\section{Analysis and Results}

\subsection{Identification and Photometry of LIFs}

\begin{figure}
    \centering
    \includegraphics[width=\linewidth]{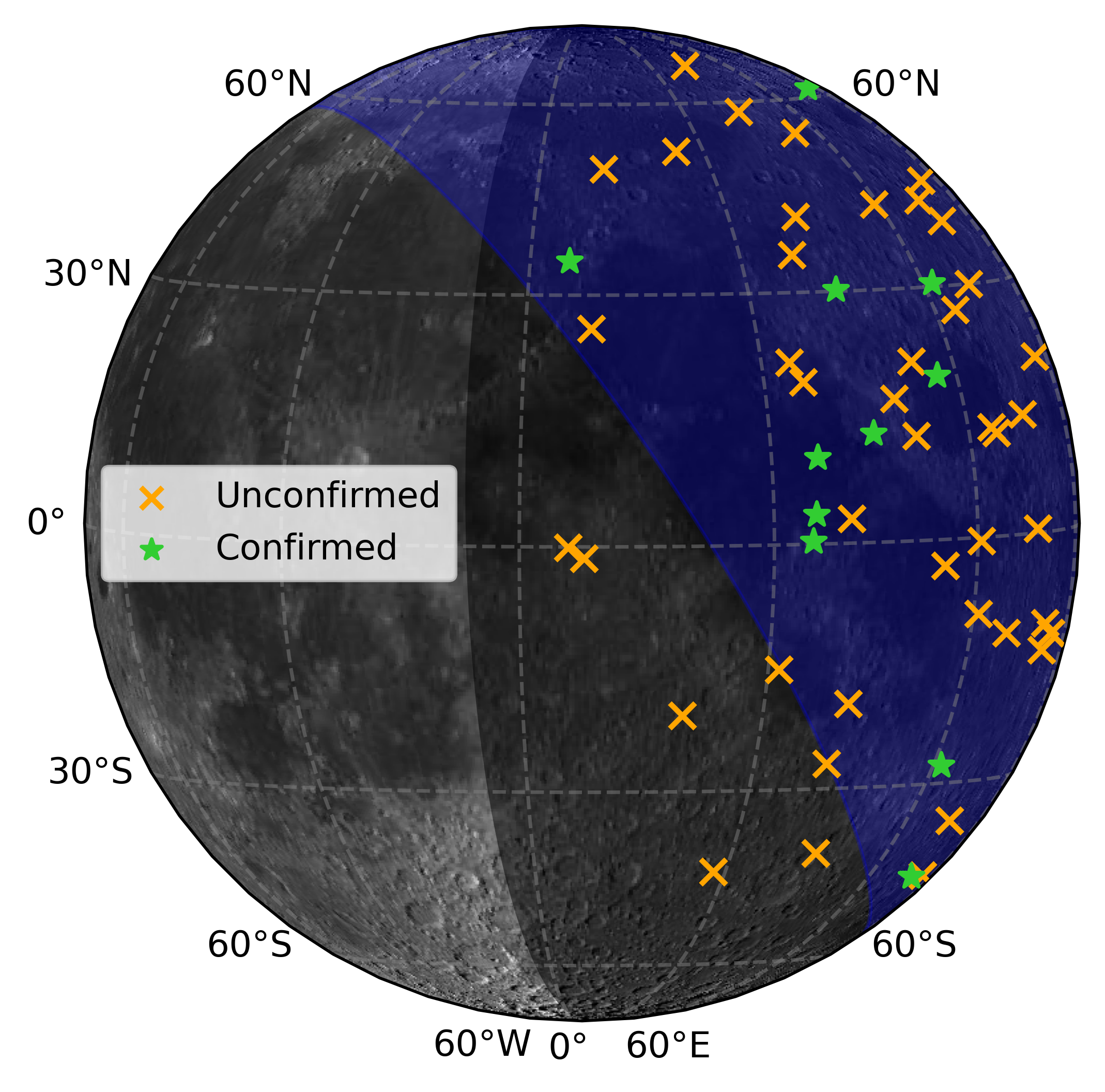}
    \caption{Illuminated hemisphere of the Moon at 2025-12-13 01:00:00 UTC, with the Geminid-seeing hemisphere of the Moon shaded blue. Orange crosses denote unconfirmed impact flashes observed by TILT1, and green crosses denote the confirmed events. Due to the observations taking place over two nights, this configuration drifts over the course of observations.}
    \label{fig:moon}
\end{figure}

The LIF detection software described and used in \citet{avdellidou2021} and \citet{sheward2024} was utilized to process the observations and identify potential LIFs. Data from each camera were examined separately. Over this course, 53 LIF candidates (visualised in Figure~\ref{fig:moon}) were detected 5\textsigma~above the noise level that could not immediately be discounted as a noise characteristic (see Table~\ref{tab:big_table}). For each of the candidate events we estimated the impact coordinates on the lunar surface using the methodology described in \citet{avdellidou2021}.
The duration of each event was calculated by adding the duration of the frames in which the LIF was detected including the readout time \citep[see for example][]{avdellidou2021}.

Due to the non-synchronisation of our camera detectors we were unable to confirm all our candidate events. Six of our events lasted for more than one frame (hereafter "multi-frame events"). Since our observations were also part of the LUMIO-issued global observing campaign to detect LIFs during the Geminids meteor shower, we utilised the candidate events from the whole team to verify our detections. Moreover, we similarly utilised the publicly available validated and candidate events from the NELIOTA programme. In total, we successfully verified eleven of our events, including the multi-frame events. Considering that the rest of the observers, excluding NELIOTA, used single detection systems with unsynchronised cameras of different types, we cannot rule out that the rest of our detections are real LIF events.

Next, we performed aperture photometry to all our detections (both confirmed and candidates) using the Source Extractor tool \citep{bertin1996} as detailed in \citet{avdellidou2019}. In order to maximise the time observing the Moon for impacts, rather than taking dedicated star calibration measurements, we instead utilised transiting stars observed within the telescopes field-of-view during lunar observations. We obtain calibrated magnitudes for the LIFs by using the observed flux from stars at similar airmass to that of the flash. We first have to calculate the extincted magnitude of the star at the observed airmass, $m_X$, using the equation

\begin{equation}\label{eq:ext}
     m_X = m_0 + \kappa X
\end{equation}

\noindent where $m_0$ is the exoatmospheric magnitude of the star, $\kappa$ is the extinction coefficient, taken as 0.25\,mag airmass$^{-1}$, and $X$ is the airmass of the star observation. Using this extincted magnitude, we can calculate the counts\,s$^{-1}$ of the star were it at the flashes airmass, using Equation~\ref{eq:ext} to similarly calculate the magnitude of the star were it at the airmass of the flash, $m_{r}$. We then perform comparative photometry using these two magnitudes and the known counts\,s$^{-1}$ of the star, to calculate the counts\,s$^{-1}$ for the star at the airmass of the flash, using the equation

\begin{equation}
    F_X = F_f \times 10^{-\frac{m_f - m_x}{2.5}}
\end{equation}

\noindent where $F_X$ is the counts\,s$^{-1}$ of the star at the airmass of the flash, and $m_f$ is the magnitude of the star at the airmass of the flash. We can then finally calculate the exoatmospheric magnitude of the observed LIF with the equation

\begin{equation}
    m_{LIF} = m_0 - 2.5\,log_{10}\left(\frac{F_{LIF}}{F_X}\right)
\end{equation}

\noindent where $F_{LIF}$ is the observed counts\,s$^{-1}$ of the LIF.

\begin{figure}
    \centering
    \includegraphics[width=\linewidth]{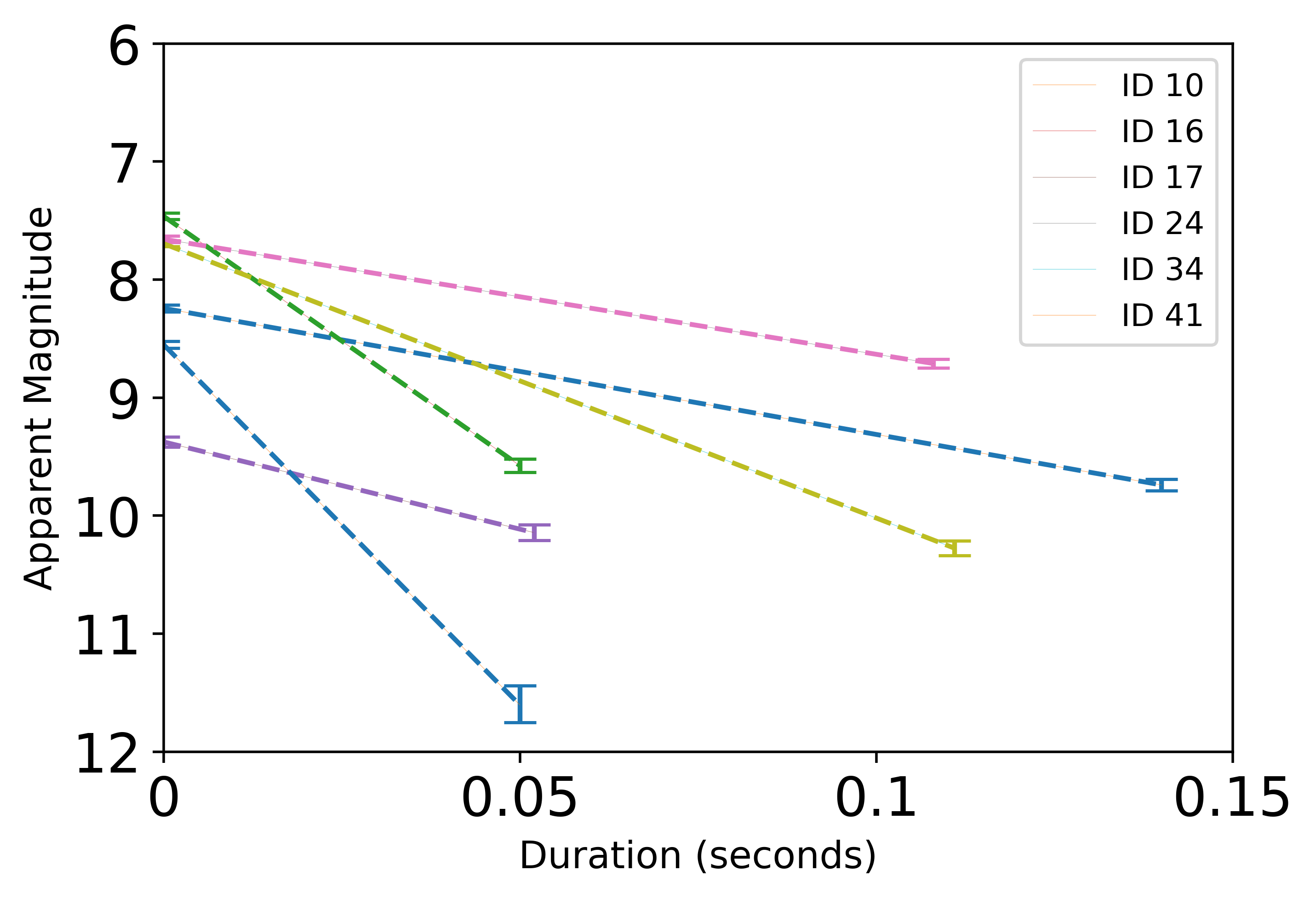}
    \caption{Magnitude decay curves for each of the six observed multi-frame events. While each event shown is only two frames long, the difference in duration is due to the computing issue, which caused large delays during readout times, or lost frames.}
    \label{fig:multiframe}
\end{figure}

All six multi-frame events each only consisted of two frames, however due to the camera-readout issues, three of these events (ID 10, 24, and 34) each had a readout times longer than 50\,msec, and thus would have been at least three frames long had the camera software correctly saved the exposures. The variable readout times, and therefore variable durations of the multi-frame events is illustrated in Figure~\ref{fig:multiframe}.

\subsection{Origin and Physical Properties of Impactors}

Although the observations were carried out around the maximum of Geminids activity, we still need to verify whether each event originated from the Geminids, from the background population, or potentially non-Geminid meteoroid streams that were active during our observations. The link (or non-link) of each impactor to a stream is a fundamental step to obtain the impact speed. Using the method described in \citet{avdellidou2019}, we found that 50 of the events in total, including all eleven confirmed events, exhibit geometry congruent with a Geminids origin (Table~\ref{tab:big_table}), leaving only 3 events which could not have a Geminid origin. It should also be noted that as the ZHR of the Geminids was 144\,h$^{-1}$, whereas the ZHR for the Monocerotids was $<3$\,h$^{-1}$, making it far more likely for an event with both as potential parent streams to be belonging to the Geminids. In order to quantify this probability, we implemented the methodology given in \cite{avdellidou2021}. We found that of the events with a possible Geminids origin, the average probability of a Geminid source was 89\%, with a minimum of 61\%.

Using the magnitudes calculated in Section 4.1, we can then calculate the luminous energy, $E_{lum}$, released by each impact. Due to the fact that the majority of the impacts were observed in only a single camera, we were unable to use the methodology given in \cite{avdellidou2021} to derive a temperature and subsequently calculate the E$_{lum}$ using Planck's law. We instead followed the methodology given in \cite{suggs2014}, to calculate the observed luminous energy of each frame the impact using the equation

\begin{equation}
    E_{lum} = 3.75\times10^{-8}\times10^{-\frac{m_{LIF}}{2.5}}\pi f\Delta\lambda D^2 t
	\label{eq:power}
\end{equation}

\noindent
where 3.75 x $10^{-8}\,W$\textmu$ $m$^{-1}$m$^{-2}$ is the flux from a magnitude 0 reference star, $m_{LIF}$ is the magnitude of the flash during that frame, $f$ is a unit-less parameter denoting the isotropy of the flash, taken here as $f$=2, denoting the light originated from the lunar surface and is radiated over half a sphere, $\Delta\lambda$ is the wavelength range of the observations, $\Delta\lambda$\,=\,0.3\,\textmu m, $D$ is the Earth-Moon distance in meters at the time of the LIF, and $t$ is the exposure time of the image frame. 

We calculate the E$_{lum}$ for each frame of the flash, and sum each frame to obtain the total $E_{lum}$ for the observed portion of multi-frame events. Due to the readout time between frames, some energy released during the LIF is not recorded, and therefore lost. In order to reconstruct an approximation of this lost energy, we used the magnitude decay slope of the observed frames to estimate the magnitude of the flash during the readout period. We can again use Equation~\ref{eq:power} to calculate the energy for the readout period using this estimated magnitude, with t taken as the readout time between the two frames. The total E$_{lum}$ of the flash is then simply the sum of the E$_{lum}$ for each frame and inter-frame period of the flash. This method of course introduces additional uncertainties to the final value. The non-observed radiant energy is not known, and over the duration of a LIF the flux of radiated energy is highly dynamic, often varying at timescales shorter than our observational setup's exposure time~\citep{giancono2026}. Consequently, the reconstructed value we obtain for the read-out period is merely a best-guess estimate, and the true value would vary dependant on the intra-frame dynamics of the LIFs evolution.

The luminous efficiency, \texteta, is the fraction of an impactor's kinetic energy which is converted into light during the impact. Within literature this value can vary across a few orders of magnitude~\citep{bellotrubio2000A, swift2011, bouley2012}, ranging from $5\times10^{-4}$ to $1.5\times10^{-3}$, and therefore can greatly affect the value for the kinetic energy obtained. For this reason, we elected to use the value for \texteta~from \citet{sheward2025}, which was calculated from the resultant craters from known LIFs. Using \texteta~we can obtain the total kinetic energy, E$_k$ with the relationship 

\begin{equation}
    E_k = \frac{E_{lum}}{\eta}
\end{equation}

\noindent
where $\eta$ is $6.0\pm1.2 \times 10^{-3}$, as measured using new techniques in \cite{sheward2025}.

By obtaining the possible parent meteoroid streams, we are able to calculate the relative velocity of the meteoroids group velocity to the velocity of the Moon, and the angle between the radiant of the meteoroid stream and the lunar surface at the impact location, i.e. the impact velocity, $V$, and impact angle, $\theta$. We can solve for the impactor mass using

\begin{equation}
    m = \frac{2E_k}{v^2}
\end{equation}

\noindent
where E$_k$ and $v$ are our calculated values. As we cannot determine for certain which stream a given impact belongs to, we obtain a velocity and mass for each of the possible parent stream cases. Similarly, even when meteoroid streams are active we cannot rule out the sporadic background population as the meteoroid source. The median probability we found for a sporadic source of our events is only 6.7\%, and is therefore unlikely to be the case compared to the much higher probability Geminids stream. Nonetheless, we must still consider this case, and thus obtain a mass by assuming an average velocity for sporadic impacts, \mbox{$v=20\,$km$\,s^{-1}$}, and assume an impact angle of $\theta=45^\circ$. 

\section{Discussion}

\begin{figure}
    \centering
    \includegraphics[width=\linewidth]{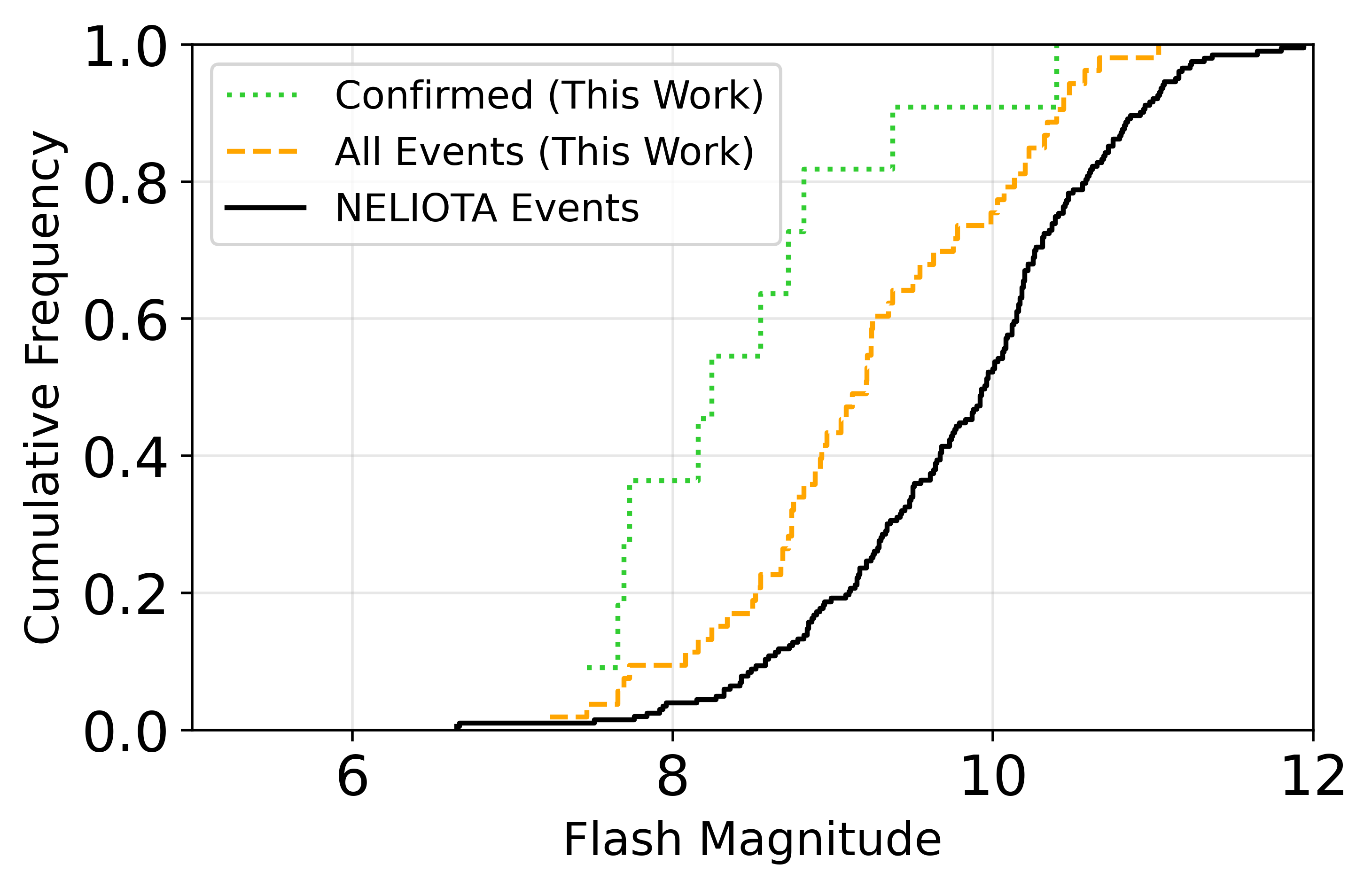}
    \caption{Relative cumulative frequency densities for confirmed LIF observations (green), and all event observations (orange), compared to that of the NELIOTA campaigns confirmed LIF observations (black).}
    \label{fig:cfd}
\end{figure}

Without secondary observations, while some false-positives can be eliminated by identifying other sources such as obvious cosmic rays, or interstitial satellites, confirmation of the single-frame events is not possible. Through previous observation campaigns such as NELIOTA, however, we can obtain the expected flash magnitude distribution of meteoroids from their observed R-band LIFs, and compare to that of our own observations. As shown in Figure~\ref{fig:cfd}, the shape of our total observations roughly matches the expected shape, albeit with less granularity, and higher frequency. The granularity is simply due to NELIOTA taking observations over 9 years, compared to our two nights. The higher frequency, one could argue, can simply indicate that some of the single-frame events are from non-LIF sources. The direct comparison of these datasets is somewhat flawed, however, due to the fact that NELIOTA observes using an R-band filter, whereas our observations were taken unfiltered. While our camera's main observational window is in the V- and R-band, some sensitivity exists up to 1000\,nm, and would therefore receive more light from a given flash.

Furthermore, NELIOTAs observations take place throughout the year, rather than targeting a specific high ZHR meteoroid stream, and therefore includes a greater proportion of sporadic meteoroid impacts than our dataset. As the Geminids meteoroid stream has a greater group velocity than the typical velocity of sporadics ($v_{GEM}$ = 35\,km\,s$^{-1}$, $v_{Spo}$ = 20\,km\,s$^{-1}$), an impactor of given size will have greater kinetic energy, and therefore produce a brighter flash. 

When considering only our confirmed LIFs we similarly obtain a higher frequency than that of NELIOTA, which indicates that the total distribution should also be higher, implying that there is the possibility of true LIFs being contained within the unconfirmed observations.
%and is therefore likely due to the aforementioned reasons, rather than the inclusion of a large number of false-positive events. 
It should be emphasised however, that this does not confirm the non-inclusion of any false-positive observations. 

Previous observations of the Geminids allow a comparison of observations rates. For observations during the peak of the Geminids, \cite{ortiz2015} obtained four LIFs in 2.4 hours, giving a rate of 1.67±0.5~LIF\,h$^{-1}$. It should be noted that the faintest flash obtained was around Mag$_v$=9.3, indicating a brighter limiting magnitude than our observational setup. When considering only confirmed LIFs with Mag$_v$=9.5 or brighter from our dataset, we obtained a rate of 1.05±0.33~LIF\,h$^{-1}$.

While \cite{yanagisawa2008} provide no total observation time data, and therefore a comparison is not possible, a similar calculation for \cite{yanagisawa2021} can be performed. \cite{yanagisawa2021} observed a total of 13 LIFs during the five hours of observations they performed during the 2018 Geminids peak, using a telescope of similar size to ours, and therefore likely had a similar limiting magnitude. This gives a rate of 2.6±0.28~LIF\,h$^{-1}$ for their observations. From our own confirmed observations, we obtained 11 LIFs in our 8.5 hour observational period, giving a rate of 1.3±0.3~LIF\,h$^{-1}$.

If we were to consider all 53 of our observations, we would obtain a much higher rate of 6.2±0.14~LIF\,h$^{-1}$. This rate is clearly in conflict with the previous observations rates of the Geminids, and indicates that there are likely several false-positive detections within the dataset. As our confirmed LIF rate is lower than the previous works, this similarly indicates that we possibly have unconfirmed positive detections within the dataset.

In order to increase our confirmed observation rate to within the error bar range of these two previous observations, we would need an extra 0.1 to 1.6 LIF\,h$^{-1}$. For our dataset, this would imply that between 1-14 of the unconfirmed impacts could be true LIFs, with the remaining 28-41 events being false-positives, potentially due to directly incident cosmic rays, or in-camera phenomena.

During this observation campaign, we were able to observe 8.5 hours, over the 48-hour period of peak Geminid activity. Since larger impacts occur at a much lower frequency, maximising observational time for the purposes of catching these 'lucky' events is crucial. A network of multiple observing stations distributed across the globe would therefore allow continuous observations (weather permitting) to increase drastically the total number of impacts detected, and thereby increase the number of higher energy impacts detected. 

These higher energy impacts are vital for seismological studies, as the seismic waves generated are proportional to the momentum of the impactor. The energy of seismic waves diminish with 1/r$^2$, and the amplitude drops with 1/r, therefore limiting the distance at which an impact is detectable above the background noise with seismometers. Consequently, maximising observational time increases the chances of both a smaller meteoroid impacting close enough to a seismometer to be detectable, and for a larger, less frequent impact to occur for which the radius of detectability is much greater. 

When considering the observation rate of NELIOTA, their most recent statistics give a rate of 3.32$\times10^{-7}$\,LIF\,h$^{-1}$\,km$^{-2}$ for confirmed stream-linked events, and 6.14$\times10^{-7}$\,LIF\,h$^{-1}$\,km$^{-2}$ for all candidate and confirmed stream events~\citep{liakos2024}. Given our telescopes field-of-view and observational configuration, we obtained an effective observation surface area of around 5.2$\times$10$^{6}$\,km$^2$, however due to the glare from the illuminated hemisphere spilling into the image, the actual surface area for which LIF detection is performed is less. The values calculated here are therefore minimum values, as the same number of events within a smaller sampling area would increase detection rates. Over the 8.5 hours of observations, we obtained a rate of 2.47$\times10^{-7}$\,LIF\,h$^{-1}$\,km$^{-2}$ for confirmed events, and 11.92$\times10^{-7}$\,LIF\,h$^{-1}$\,km$^{-2}$ for candidate events. While the confirmed LIF rate is lower (perhaps due to the overestimation of the observed area) the observation rates of candidate LIFs is considerably higher and therefore demonstrates these meteoroid streams as an invaluable opportunity for LIF observers.

It is also worth emphasizing the role of candidate LIFs. For correlating an LIF with the corresponding seismic signal of an impact, independent optical confirmation of the LIF is not strictly required. If a seismic event is detected and can be associated with an LIF, the latter is effectively confirmed by the seismometer. Conversely, if an LIF is optically confirmed but the impact is too weak or too distant to generate a detectable seismic signal, it does not contribute to the reconstruction of the lunar interior through multi-messenger (optical and seismic) observations. 

Hence, our observations and the analysis methods presented here represent the procedure that should be adopted for the optical component of the LISTEN-FLASH project. This consists of maximizing observing time, acquiring data at a high frame rate (ideally 20 frames per second or higher), and monitoring a large fraction of the lunar surface.

The observations are then analysed with a LIF detection software that produces a list of candidate events. Events that, upon human vetting, clearly do not correspond to LIFs — such as signals produced by the passage of artificial satellites, interplanetary objects, or cosmic rays, are removed, leaving a list of potential LIFs. To further reduce this list, a point-spread function (PSF) analysis can be performed to discard events whose signal above the background is incompatible with the optical PSF of the telescope (this step is not performed here).

Table~\ref{tab:big_table} shows that TILT can be sensitive to impacts of masses larger of some tens of grams. This sensitivity is compatible with most of the impacts detected by the Apollo Short Period seismometers \citep[1-30 grams impacting at distances 5-50 km][]{lognonne2009}. The sensitivity of FSS and SPSS are expected to be superior to those of the Apollo Short Period seismometers. While the events in Tab.~\ref{tab:big_table} would have occurred too far away to be detected by a seismometer at the South pole, this work presents the methods of observations relevant for linking LIFs to seismic detections.

Assuming the flux of impactors from \citet{brown2002}, after correcting for the impactors flux differences between Earth and Moon \citep{lognonne2009}, rates of 640 impactors per year are expected for masses, $m$, $\ge$1\,kg on the near side of the Moon. This rate varies as $m^{-0.9}$. This gives per year $\sim$80 impacts  $\ge$10\,kg and 10 impacts $\ge$100\,kg. This therefore highlights the necessity for long-period continuous observations for LIFs to increase the probability of recording these rarer, high mass events.

Since potential LIFs can be detected from a single station using a single camera, this work demonstrates the invaluable contribution that amateur astronomers and citizen-science projects can bring to the joint optical and seismometric observation of the Moon. To further support this concept, some of the LIFs observed by TILT were confirmed by independent observations from amateur astronomer stations operating at sites different from Calern (see Tab.~\ref{tab:big_table}), thus reinforcing the value of a global professional–amateur network of telescopes for the continuous monitoring of impacts on the Moon.

The rate of seismic impact detection on the Moon has been measured by the Apollo seismometer~\citep{duennebier1976}. This provides a rate of seismic detection given by 
\begin{equation}
    log_{10}(N)= 1.33-1.26~log_{10}(a)
\end{equation}
\noindent where $N$ is the number of impacts per year, and $a$ is the peak-to-peak seismic signal’s amplitude recorded at 5\,Hz, in nm. Typically, signals with an amplitude of 6\,DU on Apollo had an acceptable signal to noise for seismic analysis. As one DU\,=\,0.0667\,nm, this corresponds to about 0.40 nm of peak to peak amplitude and therefore a rate of about 68 impacts per year. Duennebier proposed that such signal can be detected for an impact at about 25 km, with a vertical velocity of 22.5 km\,s$^{-1}$ and an impactor mass of 10\,g. 

\begin{figure}
    \centering
    \includegraphics[width=\linewidth]{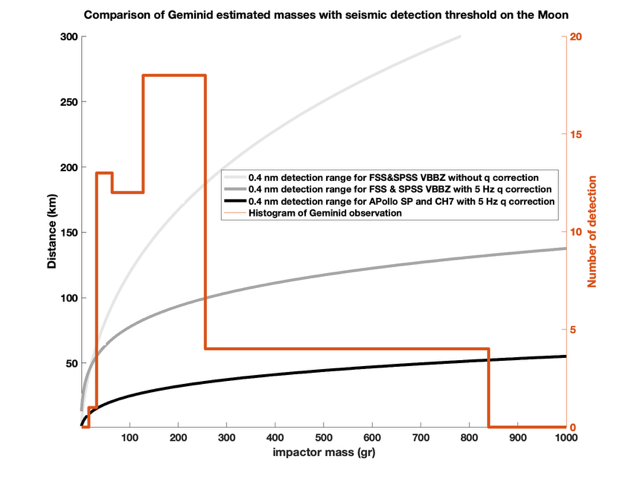}
    \caption{Comparison of the masses of the Geminid detected LIFs with the detection range of Apollo SP, Chang'e-7 broad-band, and the VBBZ instrument onboard FSS and SPSS. Impulse have been computed assuming a 20 km\,s$^{-1}$ velocity impacting at 45°.}
    \label{fig:seis}
\end{figure}

Significant trade-off remains however between the impact mass and the distance, as the seismic signal amplitude depend on the mass and distance, but also on the seismic propagation properties, including strength of attenuation and scattering, and on the impact angle and ejecta generation~\citep{lognonne2009}. Practically, this trade-off requires a fixed hypothesis on the impactor flux and impact velocities, as well as on the dependency of signal amplitude with distance, and then using an ad-hoc term, linearly tune the seismic amplitude in order to retrieve the statistical distribution. Here, we assume the impactors mass-frequency distribution as proposed by \cite{brown2002}, that the impact velocities are 20 km\,s$^{-1}$ impacting at 45°, and that the distance dependency is the one calibrated for SIV-B impacts by \cite{lognonne2009}. We also extrapolate the attenuation from 0.5\,Hz (the central frequency used by \cite{lognonne2009}, and 5\,Hz, by using a q(5\,Hz)\,=\,q(0.5\,Hz)10, where q(0.5\,Hz) is the attenuation damping factor of \cite{lognonne2009}. By tuning the linear coupling factor between acceleration and impactor mass, we retrieve the observed rate of 68 impacts per year, and have a detection threshold of about 13.6\,g  at 25\,km, slightly lower from the one extrapolated only with the velocities differences (10\,g * sqrt(2) * 22.5/20 = 16\,g). Although being empirical, this model allows to extrapolate fairly well how future seismometers can detect impacts and/or how impacts with different velocities will generate signal. With these parameter, Figure~\ref{fig:seis} shows the obtained mass distribution for the Apollo SP seismometer at 5\,Hz, compared to the distribution of impacts detected during the Geminid observations. This shows that all most of these impacts can be detected up to 30\,km, the largest one, with an estimated mass of about 840\,g, being potentially detected up to 50\,km at 5\,Hz. Note however than the latter could likely be detected at much larger distance on more long period data.

New seismometers will explore the Moon soon, such as the LS experiment onboard the Chang'e-7 mission, targeted toward Shackleton, and the FSS mission~\citep{panning2025abs}, targeted toward Schroedinger crater. At 5\,Hz, their expected resolution is 2.4$\times$10$^{-11}$m\,Hz$^{-\frac{1}{2}}$ and 8.2$\times$10$^{-13}$\,m\,Hz$^{-\frac{1}{2}}$ and almost flat in ground displacement. In a 5\,Hz bandwidth and assuming a SNR detection threshold of 3 with respect to RMS, as achieved with spectrogram techniques on the InSight HF events, the Chang’e-7 equivalent threshold corresponds to about 5 Apollo SP DU, comparable to the threshold used by \cite{duennebier1975} and to 0.165 Apollo SP DU, and therefore 36 times betters. At 5\,Hz, the detection range will be 3 time larger for VBBZ~\citep{lognonne2018}. At longer period (e.g. 0.5\,Hz), no correction is requested, which shows that impacts comparable or larger than 840\,g occurring near the lunar South Pole will be not only detected by Chang'e-7 at 5\,Hz but also by FSS at 0.4\,Hz.

\section{Conclusions}
We present results obtained during the development phase the first out of the three components of the Twin Impact Lunar telescope facilities, which are designed to constantly monitor the lunar surface for meteoroid impacts for validation of the recorded seismic activity by the upcoming deployment of lunar seismometers. Over the two peak nights (13–14 December), TILT1 operated continuously and obtained 8.5 hours of data under favourable geometry. From a preliminary analysis of these observations, 53 candidate flashes were detected and 11 were confirmed as genuine LIFs through multi-frame and multi-station validation. Not all potential LIF could be validated due to a discovery of an issue in the camera acquisition system, which was still operated in a test mode and with the cameras not synchronized between them. The confirmed flashes reached magnitudes between +7.5 and +10.4 and correspond to meteoroids of sub-kilogram scales. This campaign demonstrates that TILT is capable of autonomous LIF detection, confirmation, and photometric characterization, validating its performance for future coordinated observations with lunar seismometers. With the development of TILT2 and TILT3, the network will have at least one Moon-facing station at all times, thereby maximising observing time to maximise the chances of detecting higher energy, low frequency lunar impacts which can be detected by future lunar seismometers.

\section*{Acknowledgements}
MD, PL, P-YF acknowledge support from the A dvanced ERC grant LISTEN-FLASH. DS and CA acknowledge support from the UKSA (UKRI1377 grant: Study of the Meteoroid Impacts on the Moon).  We acknowledge support from the Bonus Qualite de Recherche (BQR) OCA and IPGP for the development of TILT1. DS also acknowledges support from the BQR of the Laboratoire Lagrange of the CNRS / OCA. MD is Leverhulme Visiting Professor at the University of Leicester with financial support from the Leverhulme Trust (UK). EMA acknowledges support by the Italian Space Agency through the agreement n. 2024-6-HH.0, "Supporto scientifico alla missione LUMIO”. LUMIO is a mission funded under ESA’s General Support Technology Programme (GSTP) through the support of the national delegations of Italy (ASI), the United Kingdom (UKSA), Norway (NOSA), and Sweden (SNSA). We thank Frederic Lambert, Christophe Limonta, Daniel Kamm, and the colleagues of the \emph{Division des Systeme Informatique} of OCA for their support.\\

%%%%%%%%%%%%%%%%%%%%%%%%%%%%%%%%%%%%%%%%%%%%%%%%%%
\section*{Data Availability}
The impact flash data used within this paper are available upon email request from the corresponding author.



\bibliographystyle{mnras}
\bibliography{references} % if your bibtex file is called example.bib

%\begin{thebibliography}{99}
%\end{thebibliography}

%%%%%%%%%%%%%%%%%%%%%%%%%%%%%%%%%%%%%%%%%%%%%%%%%%

%%%%%%%%%%%%%%%%% APPENDICES %%%%%%%%%%%%%%%%%%%%%

\appendix

\section{Table A1}

\onecolumn
\newpage
\small
\begin{landscape}
\begin{longtable}{|r|ccccccccccccc|}
\caption{List of lunar impact flash candidates observed during the Geminids. Events with multiple frames are confirmed LIFs. IDs marked with $\dagger$ denotes the event was confirmed by other observers (private communication with A. Cook, a coordinator of the LUMIO Geminids campaign). IAU stream codes are used as follows: GEM = Geminids, MON = Monocerotids, DSV = December \textsigma-Virginids, Spo = Sporadic. Stream probabilities may not sum to 100\% due to rounding.} \\
\hline\hline
ID & UTC Date & UTC Time & Lat App & Lon App & Mag & Frames & Duration & E$_k$& Stream & Probability & Velocity & Impact Angle & Impactor Mass \\
   &  &   & (Deg) & (Deg) & (Peak) & & (s) & (MJ) & Code & (\%) & (km s$^{-1}$) & (Deg) & (g) \\
\hline
\endfirsthead
\multicolumn{8}{c}
        {{ Table \thetable\ : Continued from previous page.}} \\
\hline\hline
\endhead
\hline\endfoot
1 & 2025-12-13 & 01:42:08.0 & -14.6 & 31.4 & $7.2\pm0.1$ & 1 & 0.05 & $118.8\pm2.9$ & GEM & 61.0 & 33.2 & 6.9 & $216.3\pm5.3$ \\
 &  &  &  &  &  &  &  &  & MON & 0.7 & 39.6 & 19.2 & $151.6\pm3.7$ \\
 &  &  &  &  &  &  &  &  & Spo & 38.3 & 20.0 & 45.0 & $594.2\pm14.5$ \\
\hline
2 & 2025-12-13 & 01:52:26.4 & 20.3 & 83.3 & $8.8\pm0.1$ & 1 & 0.05 & $29.2\pm1.0$ & GEM & 90.1 & 33.2 & 62.1 & $53.2\pm1.9$ \\
 &  &  &  &  &  &  &  &  & MON & 0.4 & 39.6 & 50.0 & $37.3\pm1.3$ \\
 &  &  &  &  &  &  &  &  & Spo & 9.4 & 20.0 & 45.0 & $146.2\pm5.1$ \\
\hline
3 & 2025-12-13 & 01:52:37.7 & -39.4 & 44.7 & $8.7\pm0.1$ & 1 & 0.05 & $31.5\pm1.1$ & GEM & 65.1 & 33.2 & 10.2 & $57.3\pm2.1$ \\
 &  &  &  &  &  &  &  &  & MON & 1.0 & 39.6 & 31.8 & $40.1\pm1.4$ \\
 &  &  &  &  &  &  &  &  & Spo & 33.8 & 20.0 & 45.0 & $157.3\pm5.7$ \\
\hline
4 & 2025-12-13 & 02:20:36.4 & 2.8 & 40.1 & $9.3\pm0.1$ & 1 & 0.05 & $16.9\pm0.7$ & GEM & 76.9 & 33.1 & 19.6 & $30.8\pm1.2$ \\
 &  &  &  &  &  &  &  &  & MON & 0.5 & 39.6 & 24.1 & $21.6\pm0.8$ \\
 &  &  &  &  &  &  &  &  & Spo & 22.5 & 20.0 & 45.0 & $84.6\pm3.3$ \\
\hline
5 & 2025-12-13 & 03:06:13.4 & 41.7 & 72.2 & $9.1\pm0.1$ & 1 & 0.05 & $21.6\pm0.7$ & GEM & 88.4 & 33.1 & 45.3 & $39.4\pm1.2$ \\
 &  &  &  &  &  &  &  &  & MON & 0.3 & 39.6 & 27.3 & $27.6\pm0.9$ \\
 &  &  &  &  &  &  &  &  & Spo & 11.3 & 20.0 & 45.0 & $108.0\pm3.4$ \\
\hline
6 & 2025-12-13 & 03:40:10.1 & 44.4 & 79.8 & $8.5\pm0.1$ & 1 & 0.05 & $36.9\pm0.9$ & GEM & 90.0 & 33.1 & 48.3 & $67.4\pm1.7$ \\
 &  &  &  &  &  &  &  &  & MON & 0.3 & 39.5 & 27.9 & $47.3\pm1.2$ \\
 &  &  &  &  &  &  &  &  & Spo & 9.7 & 20.0 & 45.0 & $184.7\pm4.7$ \\
\hline
7 & 2025-12-13 & 03:43:27.0 & -1.3 & 7.5 & $10.6\pm0.1$ & 1 & 0.05 & $5.5\pm0.4$ & Spo & 100.0 & 20.0 & 45.0 & $27.3\pm2.0$ \\
\hline
8 & 2025-12-13 & 04:30:31.0 & -20.1 & 19.6 & $9.2\pm0.1$ & 1 & 0.05 & $19.2\pm0.6$ & MON & 1.0 & 39.5 & 10.4 & $24.6\pm0.8$ \\
 &  &  &  &  &  &  &  &  & Spo & 99.0 & 20.0 & 45.0 & $96.0\pm3.2$ \\
\hline
9 & 2025-12-13 & 04:38:28.4 & -8.9 & 61.0 & $9.2\pm0.1$ & 1 & 0.05 & $19.2\pm0.6$ & GEM & 87.6 & 33.1 & 37.5 & $34.9\pm1.1$ \\
 &  &  &  &  &  &  &  &  & MON & 0.5 & 39.5 & 48.3 & $24.6\pm0.8$ \\
 &  &  &  &  &  &  &  &  & Spo & 11.9 & 20.0 & 45.0 & $95.8\pm3.0$ \\
\hline
10 & 2025-12-13 & 05:36:32.8 & 0.5 & 73.7 & $9.5\pm0.1$ & 1 & 0.05 & $14.7\pm0.5$ & GEM & 90.8 & 33.2 & 53.2 & $26.7\pm0.9$ \\
 &  &  &  &  &  &  &  &  & MON & 0.4 & 39.5 & 57.9 & $18.9\pm0.6$ \\
 &  &  &  &  &  &  &  &  & Spo & 8.8 & 20.0 & 45.0 & $73.4\pm2.4$ \\
\hline
11 & 2025-12-13 & 05:48:57.9 & 30.2 & 43.4 & $8.2\pm0.1$ & 2 & 0.19 & $100.7\pm3.3$ & GEM & 88.7 & 33.1 & 26.9 & $183.3\pm6.1$ \\
 &  &  &  &  &  &  &  &  & MON & 0.2 & 39.5 & 18.0 & $129.2\pm4.3$ \\
 &  &  &  &  &  &  &  &  & Spo & 11.1 & 20.0 & 45.0 & $503.3\pm16.7$ \\
\hline
12 & 2025-12-13 & 06:02:22.9 & -10.7 & 78.6 & $11.0\pm0.1$ & 1 & 0.05 & $3.6\pm0.2$ & GEM & 88.9 & 33.1 & 52.5 & $6.5\pm0.3$ \\
 &  &  &  &  &  &  &  &  & MON & 0.5 & 39.5 & 66.4 & $4.6\pm0.2$ \\
 &  &  &  &  &  &  &  &  & Spo & 10.5 & 20.0 & 45.0 & $17.9\pm0.9$ \\
\hline
13 & 2025-12-14 & 02:35:18.7 & 21.3 & 33.8 & $8.5\pm0.1$ & 1 & 0.05 & $36.3\pm1.5$ & GEM & 92.3 & 32.9 & 27.7 & $67.2\pm2.7$ \\
 &  &  &  &  &  &  &  &  & MON & 0.1 & 39.3 & 22.9 & $46.9\pm1.9$ \\
 &  &  &  &  &  &  &  &  & Spo & 7.6 & 20.0 & 45.0 & $181.2\pm7.3$ \\
\hline
14 & 2025-12-14 & 02:41:52.1 & -35.6 & 72.6 & $9.2\pm0.1$ & 1 & 0.05 & $18.6\pm0.7$ & GEM & 93.1 & 32.8 & 36.8 & $34.5\pm1.3$ \\
 &  &  &  &  &  &  &  &  & MON & 0.2 & 39.3 & 62.0 & $24.1\pm0.9$ \\
 &  &  &  &  &  &  &  &  & Spo & 6.7 & 20.0 & 45.0 & $92.9\pm3.6$ \\
\hline
15 & 2025-12-14 & 02:49:10.7 & 35.0 & 38.2 & $9.2\pm0.1$ & 1 & 0.05 & $19.1\pm0.7$ & GEM & 92.3 & 32.8 & 31.1 & $35.4\pm1.3$ \\
 &  &  &  &  &  &  &  &  & MON & 0.1 & 39.3 & 19.5 & $24.7\pm0.9$ \\
 &  &  &  &  &  &  &  &  & Spo & 7.6 & 20.0 & 45.0 & $95.3\pm3.5$ \\
\hline
16 & 2025-12-14 & 03:04:26.1 & 58.0 & 43.7 & $9.8\pm0.1$ & 1 & 0.05 & $11.6\pm0.5$ & GEM & 93.6 & 33.0 & 28.3 & $21.4\pm0.9$ \\
 &  &  &  &  &  &  &  &  & MON & 0.0 & 39.3 & 8.2 & $15.0\pm0.6$ \\
 &  &  &  &  &  &  &  &  & Spo & 6.4 & 20.0 & 45.0 & $58.1\pm2.3$ \\
\hline
17 & 2025-12-14 & 03:09:10.7 & 3.4 & 35.4 & $7.5\pm0.1$ & 2 & 0.10 & $109.4\pm3.1$ & GEM & 95.2 & 33.0 & 27.4 & $201.2\pm5.7$ \\
 &  &  &  &  &  &  &  &  & MON & 0.1 & 39.3 & 31.5 & $141.6\pm4.0$ \\
 &  &  &  &  &  &  &  &  & Spo & 4.7 & 20.0 & 45.0 & $546.8\pm15.5$ \\
\hline
18 & 2025-12-14 & 03:11:33.4 & 12.6 & 44.1 & $9.4\pm0.1$ & 2 & 0.10 & $25.0\pm1.2$ & GEM & 95.3 & 33.0 & 37.3 & $46.0\pm2.2$ \\
 &  &  &  &  &  &  &  &  & MON & 0.1 & 39.3 & 35.9 & $32.4\pm1.5$ \\
 &  &  &  &  &  &  &  &  & Spo & 4.6 & 20.0 & 45.0 & $125.0\pm5.9$ \\
\hline
19 & 2025-12-14 & 03:15:08.2 & 12.1 & 50.7 & $8.7\pm0.1$ & 1 & 0.05 & $31.0\pm0.8$ & GEM & 96.0 & 33.0 & 43.8 & $57.0\pm1.5$ \\
 &  &  &  &  &  &  &  &  & MON & 0.1 & 39.3 & 41.8 & $40.1\pm1.1$ \\
 &  &  &  &  &  &  &  &  & Spo & 3.9 & 20.0 & 45.0 & $154.9\pm4.1$ \\
\hline
20 & 2025-12-14 & 03:16:05.8 & -0.1 & 5.6 & $8.7\pm0.1$ & 1 & 0.05 & $29.4\pm0.8$ & MON & 0.2 & 39.3 & 3.8 & $38.1\pm1.1$ \\
 &  &  &  &  &  &  &  &  & Spo & 99.8 & 20.0 & 45.0 & $147.2\pm4.2$ \\
\hline
21 & 2025-12-14 & 03:19:15.2 & 38.5 & 74.8 & $10.1\pm0.1$ & 1 & 0.05 & $8.2\pm0.4$ & GEM & 95.9 & 33.0 & 56.2 & $15.0\pm0.6$ \\
 &  &  &  &  &  &  &  &  & MON & 0.1 & 39.3 & 35.4 & $10.6\pm0.5$ \\
 &  &  &  &  &  &  &  &  & Spo & 4.0 & 20.0 & 45.0 & $40.9\pm1.8$ \\
\hline
22 & 2025-12-14 & 03:33:30.8 & -11.4 & 67.7 & $9.1\pm0.1$ & 1 & 0.05 & $20.8\pm0.6$ & GEM & 95.5 & 32.9 & 52.2 & $38.4\pm1.1$ \\
 &  &  &  &  &  &  &  &  & MON & 0.1 & 39.3 & 66.6 & $26.9\pm0.8$ \\
 &  &  &  &  &  &  &  &  & Spo & 4.3 & 20.0 & 45.0 & $103.8\pm2.9$ \\
\hline
23 & 2025-12-14 & 03:58:15.0 & -0.4 & 60.7 & $9.2\pm0.1$ & 1 & 0.05 & $18.5\pm0.7$ & GEM & 95.2 & 32.9 & 51.4 & $34.2\pm1.3$ \\
 &  &  &  &  &  &  &  &  & MON & 0.1 & 39.3 & 56.8 & $24.0\pm0.9$ \\
 &  &  &  &  &  &  &  &  & Spo & 4.7 & 20.0 & 45.0 & $92.3\pm3.4$ \\
\hline
24 & 2025-12-14 & 04:00:06.2 & -12.0 & 81.7 & $9.6\pm0.1$ & 1 & 0.05 & $13.0\pm0.4$ & GEM & 95.4 & 32.9 & 61.5 & $24.1\pm0.7$ \\
 &  &  &  &  &  &  &  &  & MON & 0.1 & 39.3 & 80.4 & $16.9\pm0.5$ \\
 &  &  &  &  &  &  &  &  & Spo & 4.4 & 20.0 & 45.0 & $65.0\pm2.0$ \\
\hline
25 & 2025-12-14 & 04:15:02.0 & -27.5 & 61.7 & $7.7\pm0.1$ & 2 & 0.16 & $167.4\pm4.9$ & GEM & 95.5 & 32.9 & 37.6 & $310.1\pm9.1$ \\
 &  &  &  &  &  &  &  &  & MON & 0.1 & 39.3 & 59.2 & $217.3\pm6.4$ \\
 &  &  &  &  &  &  &  &  & Spo & 4.4 & 20.0 & 45.0 & $837.1\pm24.5$ \\
\hline
26 & 2025-12-14 & 04:15:41.2 & 48.1 & 11.0 & $8.9\pm0.1$ & 1 & 0.05 & $25.7\pm0.7$ & GEM & 83.6 & 32.9 & 11.0 & $47.6\pm1.3$ \\
 &  &  &  &  &  &  &  &  & Spo & 16.4 & 20.0 & 45.0 & $128.4\pm3.4$ \\
\hline
27 & 2025-12-14 & 04:27:59.8 & 53.6 & 53.4 & $9.0\pm0.1$ & 1 & 0.05 & $24.0\pm0.6$ & GEM & 93.8 & 32.9 & 35.5 & $44.5\pm1.2$ \\
 &  &  &  &  &  &  &  &  & MON & 0.1 & 39.3 & 15.4 & $31.2\pm0.8$ \\
 &  &  &  &  &  &  &  &  & Spo & 6.2 & 20.0 & 45.0 & $120.2\pm3.2$ \\
\hline
28 & 2025-12-14 & 04:29:39.9 & -41.9 & 28.0 & $10.1\pm0.1$ & 1 & 0.05 & $8.7\pm0.3$ & GEM & 72.6 & 32.9 & 7.0 & $16.1\pm0.5$ \\
 &  &  &  &  &  &  &  &  & MON & 0.4 & 39.3 & 29.0 & $11.3\pm0.4$ \\
 &  &  &  &  &  &  &  &  & Spo & 27.0 & 20.0 & 45.0 & $43.4\pm1.4$ \\
\hline
29 & 2025-12-14 & 04:34:40.5 & 41.6 & 59.0 & $10.7\pm0.1$ & 1 & 0.05 & $5.0\pm0.2$ & GEM & 93.4 & 32.9 & 45.1 & $9.3\pm0.4$ \\
 &  &  &  &  &  &  &  &  & MON & 0.1 & 39.2 & 27.3 & $6.5\pm0.3$ \\
 &  &  &  &  &  &  &  &  & Spo & 6.5 & 20.0 & 45.0 & $25.1\pm1.0$ \\
\hline
30 & 2025-12-14 & 04:34:55.9 & -44.3 & 80.3 & $8.1\pm0.1$ & 1 & 0.05 & $54.1\pm1.4$ & GEM & 93.8 & 32.9 & 31.9 & $100.3\pm2.5$ \\
 &  &  &  &  &  &  &  &  & MON & 0.2 & 39.2 & 58.1 & $70.3\pm1.8$ \\
 &  &  &  &  &  &  &  &  & Spo & 6.0 & 20.0 & 45.0 & $270.7\pm6.8$ \\
\hline
31 & 2025-12-14 & 04:35:05.0 & 40.3 & 41.5 & $8.9\pm0.1$ & 1 & 0.05 & $25.0\pm0.8$ & GEM & 93.4 & 32.9 & 33.4 & $46.3\pm1.5$ \\
 &  &  &  &  &  &  &  &  & MON & 0.1 & 39.2 & 19.1 & $32.4\pm1.1$ \\
 &  &  &  &  &  &  &  &  & Spo & 6.5 & 20.0 & 45.0 & $124.8\pm4.1$ \\
\hline
32 & 2025-12-14 & 04:36:10.0 & 20.9 & 52.3 & $10.0\pm0.1$ & 1 & 0.05 & $9.0\pm0.3$ & GEM & 94.0 & 32.9 & 46.0 & $16.7\pm0.6$ \\
 &  &  &  &  &  &  &  &  & MON & 0.1 & 39.2 & 38.5 & $11.7\pm0.4$ \\
 &  &  &  &  &  &  &  &  & Spo & 5.9 & 20.0 & 45.0 & $45.1\pm1.6$ \\
\hline
33 & 2025-12-14 & 04:36:53.1 & 18.9 & 35.3 & $10.4\pm0.1$ & 1 & 0.05 & $6.2\pm0.3$ & GEM & 91.1 & 32.9 & 30.0 & $11.4\pm0.5$ \\
 &  &  &  &  &  &  &  &  & MON & 0.1 & 39.2 & 26.1 & $8.0\pm0.3$ \\
 &  &  &  &  &  &  &  &  & Spo & 8.7 & 20.0 & 45.0 & $30.8\pm1.3$ \\
\hline
34$\dagger$ & 2025-12-14 & 04:36:57.3 & 34.5 & 5.5 & $8.7\pm0.1$ & 1 & 0.05 & $30.0\pm0.9$ & GEM & 70.4 & 32.9 & 5.1 & $55.6\pm1.7$ \\
 &  &  &  &  &  &  &  &  & Spo & 29.6 & 20.0 & 45.0 & $149.9\pm4.5$ \\
\hline
35 & 2025-12-14 & 04:38:05.9 & 0.3 & 35.0 & $7.7\pm0.1$ & 2 & 0.16 & $113.1\pm3.4$ & GEM & 93.6 & 32.9 & 27.1 & $209.7\pm6.3$ \\
 &  &  &  &  &  &  &  &  & MON & 0.1 & 39.2 & 32.8 & $147.0\pm4.4$ \\
 &  &  &  &  &  &  &  &  & Spo & 6.3 & 20.0 & 45.0 & $565.7\pm16.9$ \\
\hline
36 & 2025-12-14 & 04:42:25.3 & 30.0 & 71.0 & $9.5\pm0.1$ & 1 & 0.05 & $14.1\pm0.4$ & GEM & 95.2 & 32.9 & 59.5 & $26.1\pm0.8$ \\
 &  &  &  &  &  &  &  &  & MON & 0.1 & 39.2 & 42.2 & $18.3\pm0.5$ \\
 &  &  &  &  &  &  &  &  & Spo & 4.7 & 20.0 & 45.0 & $70.5\pm2.1$ \\
\hline
37 & 2025-12-14 & 04:50:25.8 & -26.6 & 40.6 & $10.3\pm0.1$ & 1 & 0.05 & $6.9\pm0.3$ & GEM & 88.9 & 32.9 & 23.0 & $12.7\pm0.5$ \\
 &  &  &  &  &  &  &  &  & MON & 0.2 & 39.2 & 41.3 & $8.9\pm0.4$ \\
 &  &  &  &  &  &  &  &  & Spo & 10.9 & 20.0 & 45.0 & $34.3\pm1.4$ \\
\hline
38 & 2025-12-14 & 05:08:17.4 & 69.1 & 42.7 & $9.1\pm0.1$ & 1 & 0.05 & $22.2\pm0.6$ & GEM & 90.6 & 32.8 & 23.1 & $41.1\pm1.0$ \\
 &  &  &  &  &  &  &  &  & Spo & 9.4 & 20.0 & 45.0 & $110.9\pm2.8$ \\
\hline
39 & 2025-12-14 & 05:09:23.8 & 12.6 & 64.7 & $10.2\pm0.1$ & 1 & 0.05 & $7.5\pm0.2$ & GEM & 94.5 & 32.8 & 58.4 & $13.9\pm0.4$ \\
 &  &  &  &  &  &  &  &  & MON & 0.1 & 39.2 & 53.1 & $9.8\pm0.3$ \\
 &  &  &  &  &  &  &  &  & Spo & 5.3 & 20.0 & 45.0 & $37.6\pm1.2$ \\
\hline
40 & 2025-12-14 & 05:09:45.1 & 16.6 & 48.1 & $10.3\pm0.1$ & 1 & 0.05 & $6.8\pm0.2$ & GEM & 93.1 & 32.8 & 42.4 & $12.5\pm0.4$ \\
 &  &  &  &  &  &  &  &  & MON & 0.1 & 39.2 & 38.1 & $8.8\pm0.3$ \\
 &  &  &  &  &  &  &  &  & Spo & 6.8 & 20.0 & 45.0 & $33.8\pm1.2$ \\
\hline
41 & 2025-12-14 & 05:09:55.2 & 11.9 & 65.6 & $10.0\pm0.1$ & 1 & 0.05 & $9.4\pm0.3$ & GEM & 94.7 & 32.8 & 59.2 & $17.4\pm0.5$ \\
 &  &  &  &  &  &  &  &  & MON & 0.1 & 39.2 & 54.2 & $12.2\pm0.4$ \\
 &  &  &  &  &  &  &  &  & Spo & 5.1 & 20.0 & 45.0 & $46.8\pm1.4$ \\
\hline
42 & 2025-12-14 & 05:11:39.3 & -44.2 & 74.7 & $8.6\pm0.1$ & 2 & 0.10 & $37.2\pm1.3$ & GEM & 92.9 & 32.8 & 30.5 & $69.1\pm2.4$ \\
 &  &  &  &  &  &  &  &  & MON & 0.2 & 39.2 & 56.6 & $48.4\pm1.6$ \\
 &  &  &  &  &  &  &  &  & Spo & 6.9 & 20.0 & 45.0 & $186.2\pm6.3$ \\
\hline
43 & 2025-12-14 & 05:16:24.2 & 50.9 & 24.7 & $8.3\pm0.1$ & 1 & 0.05 & $42.6\pm1.0$ & GEM & 91.3 & 32.9 & 20.2 & $78.8\pm1.9$ \\
 &  &  &  &  &  &  &  &  & MON & 0.0 & 39.2 & 3.7 & $55.4\pm1.3$ \\
 &  &  &  &  &  &  &  &  & Spo & 8.7 & 20.0 & 45.0 & $213.1\pm5.0$ \\
\hline
44 & 2025-12-14 & 05:23:18.4 & -13.9 & 79.3 & $9.2\pm0.1$ & 1 & 0.05 & $18.6\pm0.5$ & GEM & 95.7 & 32.9 & 59.0 & $34.4\pm0.9$ \\
 &  &  &  &  &  &  &  &  & MON & 0.1 & 39.2 & 78.9 & $24.2\pm0.7$ \\
 &  &  &  &  &  &  &  &  & Spo & 4.2 & 20.0 & 45.0 & $93.1\pm2.5$ \\
\hline
45$\dagger$ & 2025-12-14 & 05:25:31.9 & 19.1 & 56.3 & $8.8\pm0.1$ & 1 & 0.05 & $27.4\pm0.7$ & GEM & 95.5 & 32.9 & 50.2 & $50.7\pm1.3$ \\
 &  &  &  &  &  &  &  &  & MON & 0.1 & 39.2 & 42.9 & $35.6\pm0.9$ \\
 &  &  &  &  &  &  &  &  & Spo & 4.4 & 20.0 & 45.0 & $137.1\pm3.6$ \\
\hline
46$\dagger$ & 2025-12-14 & 05:32:33.1 & 10.0 & 36.0 & $8.2\pm0.1$ & 1 & 0.05 & $50.4\pm1.3$ & GEM & 93.8 & 32.9 & 30.3 & $93.2\pm2.3$ \\
 &  &  &  &  &  &  &  &  & MON & 0.1 & 39.2 & 30.9 & $65.5\pm1.6$ \\
 &  &  &  &  &  &  &  &  & Spo & 6.1 & 20.0 & 45.0 & $251.9\pm6.3$ \\
\hline
47 & 2025-12-14 & 05:36:12.7 & 13.8 & 73.0 & $8.7\pm0.1$ & 1 & 0.05 & $29.4\pm0.7$ & GEM & 96.2 & 32.9 & 66.7 & $54.4\pm1.3$ \\
 &  &  &  &  &  &  &  &  & MON & 0.1 & 39.2 & 57.4 & $38.2\pm0.9$ \\
 &  &  &  &  &  &  &  &  & Spo & 3.7 & 20.0 & 45.0 & $147.0\pm3.5$ \\
\hline
48 & 2025-12-14 & 05:36:26.0 & 25.7 & 8.4 & $8.9\pm0.1$ & 1 & 0.05 & $24.8\pm0.6$ & GEM & 75.3 & 32.9 & 6.4 & $45.8\pm1.1$ \\
 &  &  &  &  &  &  &  &  & Spo & 24.7 & 20.0 & 45.0 & $123.9\pm3.0$ \\
\hline
49$\dagger$ & 2025-12-14 & 05:39:01.1 & 30.5 & 62.0 & $10.4\pm0.1$ & 1 & 0.05 & $6.4\pm0.3$ & GEM & 94.5 & 32.9 & 52.7 & $11.9\pm0.6$ \\
 &  &  &  &  &  &  &  &  & MON & 0.1 & 39.2 & 37.9 & $8.3\pm0.4$ \\
 &  &  &  &  &  &  &  &  & Spo & 5.4 & 20.0 & 45.0 & $32.1\pm1.7$ \\
\hline
50 & 2025-12-14 & 05:45:08.8 & -3.0 & 54.2 & $10.2\pm0.1$ & 1 & 0.05 & $7.7\pm0.4$ & GEM & 93.9 & 32.9 & 45.2 & $14.2\pm0.7$ \\
 &  &  &  &  &  &  &  &  & MON & 0.1 & 39.2 & 52.6 & $10.0\pm0.5$ \\
 &  &  &  &  &  &  &  &  & Spo & 5.9 & 20.0 & 45.0 & $38.4\pm1.8$ \\
\hline
51 & 2025-12-14 & 06:04:52.7 & -19.0 & 41.7 & $9.8\pm0.1$ & 1 & 0.05 & $11.3\pm0.3$ & GEM & 91.4 & 32.9 & 27.8 & $21.0\pm0.6$ \\
 &  &  &  &  &  &  &  &  & MON & 0.2 & 39.2 & 43.2 & $14.7\pm0.4$ \\
 &  &  &  &  &  &  &  &  & Spo & 8.4 & 20.0 & 45.0 & $56.7\pm1.6$ \\
\hline
52$\dagger$ & 2025-12-14 & 06:14:10.0 & 61.9 & 82.0 & $7.7\pm0.1$ & 1 & 0.05 & $74.8\pm1.8$ & GEM & 79.8 & 32.8 & 38.7 & $138.8\pm3.3$ \\
 &  &  &  &  &  &  &  &  & MON & 0.0 & 39.2 & 14.1 & $97.4\pm2.3$ \\
 &  &  &  &  &  &  &  &  & DSV & 15.7 & 66.3 & 5.7 & $34.0\pm0.8$ \\
 &  &  &  &  &  &  &  &  & Spo & 4.5 & 20.0 & 45.0 & $374.2\pm8.8$ \\
\hline
53 & 2025-12-14 & 06:19:28.3 & 26.9 & 64.5 & $10.5\pm0.1$ & 1 & 0.05 & $6.0\pm0.2$ & GEM & 94.3 & 32.9 & 56.4 & $11.0\pm0.4$ \\
 &  &  &  &  &  &  &  &  & MON & 0.1 & 39.2 & 42.3 & $7.7\pm0.3$ \\
 &  &  &  &  &  &  &  &  & Spo & 5.6 & 20.0 & 45.0 & $29.8\pm1.2$ 
\label{tab:big_table}
\end{longtable}
\end{landscape}
\twocolumn
%%%%%%%%%%%%%%%%%%%%%%%%%%%%%%%%%%%%%%%%%%%%%%%%%%

% Don't change these lines
\bsp	% typesetting comment
\label{lastpage}
\end{document}